\documentclass[aps,twocolumn,pra]{revtex4-1}
\usepackage{graphics}
\usepackage{dcolumn}% Align table columns on decimal point
\usepackage{bm}% bold math
\usepackage{amsmath}
\usepackage{hyperref}
\hypersetup{colorlinks=true, urlcolor=blue}
\usepackage{color}% color to the text 
\usepackage{tikz}
\usepackage{multirow}

\usepackage{float}
\usepackage[caption = false]{subfig}
\usepackage{newlfont}
\usepackage{amssymb}
\usepackage{amsfonts}
\usepackage{amsthm}
\usepackage{graphicx}
\usepackage{bm}

\def \be{\begin{equation}}
\def \ee{ \end{equation} }

\begin{document}

\definecolor{red}{rgb}{1,0,0}
%\title{Device-independent Secure QKD via Informational Complementarity}
\title{Information complementarity in multipartite  quantum states\\ and security in cryptography}
%\title{Secure Communication vs Low-dimensional Eves and/or Large-dimensional Quantum channels}
%\title{}
\author{Anindita Bera\(^{1,2}\), Asutosh Kumar\(^{2}\), Debraj Rakshit\(^{2}\), R. Prabhu\(^{2}\), Aditi Sen(De)\(^{2}\), and Ujjwal Sen\(^{2}\)}
%\author{Anindita Bera, Asutosh Kumar, Aditi Sen (De), and Ujjwal Sen}
%\email{asukumar@hri.res.in}

\affiliation{\(^1\)Department of Applied Mathematics, University of Calcutta, 92 Acharya Prafulla Chandra Road, Kolkata 700 009, India\\
\(^2\)Harish-Chandra Research Institute, Chhatnag Road, Jhunsi, Allahabad 211 019, India}

\begin{abstract}
We derive complementarity relations for arbitrary quantum states of multiparty systems, of arbitrary number of parties and dimensions, 
between the purity of a part of the system and 
several correlation quantities, including entanglement and other quantum correlations as well as classical and total correlations,
of that part with the remainder of the system.
We subsequently use such a complementarity relation, between purity and quantum mutual information in the tripartite scenario,
to 
provide a bound on the 
secret key rate
for individual attacks on a quantum key distribution protocol.
%We discuss the implications of the result for the possibility of device independent security proofs of quantum cryptography without resorting to Bell inequality violations.
\end{abstract}

\maketitle

%{\it Introduction}.--
\section{Introduction}
%%%%%%%%%%%%%%%%%%%%%%
Quantum key distribution (QKD) is a protocol that allows two distant parties to share a secret without meeting and in the presence of a malicious eavesdropper
\cite{cryptoreview}. The security of the protocol is based on the validity of quantum mechanics \cite{Shor-Preskill}. 
%
% Security proofs have also been 
% considered for  eavesdroppers with supra-quantum powers \cite{supra-quantum}. An interesting and important security proof considers the case when the devices used by the 
% legitimate users of the key distribution channel cannot be trusted - the device independent scenario \cite{ekert91, mayers-yao, acin-diqkd, diqkd-NJP}. The device 
% independent security proofs rely on the violation of Bell inequality \cite{Bell} by the shared state, and   
% %The device independent security proofs 
% are vulnerable to the loopholes \cite{loopholes} in experiments that violate Bell inequalities (see \cite{diqkd-NJP} however). 
% It is therefore relevant to look for device independent security proofs that do not directly use the violation of a Bell inequality. 
%
Quantum key distribution protocols broadly fall under two main categories, viz. the product-state and the entanglement based ones, the Bennett-Brassard 1984 (BB84)
\cite{bb84} and the Ekert 1991 \cite{ekert91} protocols being prominent examples of the respective categories. It was later realized that the two categories 
are similar from several perspectives \cite{curty}, with a notable difference being in the 
%The Ekert protocol 
%of quantum  
%The 
device independent security proof \cite{ekert91, mayers-yao, acin-diqkd, diqkd-NJP}, which is based on the Ekert protocol with security obtained 
via
%. The security of the 
%Ekert protocol is usually based on the 
violation of Bell inequality \cite{Bell}. 
%While just violation is sufficient \cite{acin-diqkd, diqkd-NJP}, 
We may note that in the two-qubit scenario, 
strong violation of 
Bell inequality is possible only by states close to maximally entangled states, and such states are almost pure \cite{Garisto-Hardy, Garisto-Hardy-1}. An almost pure state 
shared between the legitimate users of the key distribution channel implies that these users are informationally detached from the rest of the world, including a 
possible eavesdropper, indicating the security of the information flowing in the channel between the legitimate users. 
%The security proofs of the Ekert protocol start off from the violation of Bell inequality by the state shared by the the legitimate users. 

In this work, we ask whether we can start a step later to obtain a security proof of quantum cryptography for individual attacks via a variation \cite{acin-protocol} 
of the Ekert key distribution protocol. Precisely, we consider states with high purity and that also has correlations which allow generation of 
correlated bit sequences. We consider a scenario with three parties called Alice, Bob, and Eve, where Alice and Bob represent the legitimate users of the protocol, 
while Eve represents the potential eavesdropper. We prove complementarity relations between the purity of the 
%legitimate users
Alice-Bob system and the quantum mutual information in the 
Alice-Bob versus Eve bipartition.
%, where we have, as is usual, christened the legitimate users as Alice and Bob, and the eavesdropper as Eve. 
We then show that such a complementarity can potentially lead to a bound on the secret key rate 
%security proof 
in the case of individual attacks.
%, implying that the security proof 
%of the Ekert protocol can be 
This secret key rate is obtained by concepts independent of Bell inequalities, which, along with providing another perspective of the 
security of the Ekert protocol, 
can also be important to avoid vulnerability of the proof from loopholes in experiments that violate Bell inequalities \cite{loopholes, diqkd-NJP} 
(cf. \cite{natun-loopholes-free}).
Note that the complementarity relation is true for any quantum state of arbitrary dimensions of the individual parties, and is 
also true for an arbitrary number of parties, and so can have applications in other quantum information protocols with or without security.

On the way, we also show that a similar complementarity exists in all multiparty quantum systems between the
purity of the ``legitimate'' users, and a large number of quantum characteristics including quantum correlations in the 
legitimate versus ``eavesdroppers'' bipartition, irrespective of whether the multipartite quantum state is pure or mixed and irrespective of the dimensions of the 
subsystems and the number of parties involved. 
We also numerically investigate the tightness of the obtained inequalities by Haar uniformly generating states of three qubits of different ranks. 

The paper is organized as follows. In the following section, 
we derive the complementarity relations. Their tightness is considered in the Sec. \ref{tightness}. In the 
succeeding section (Sec. \ref{sec-crypto}), 
we use a complementarity relation to provide bounds on the secret key rate. A concluding section is presented at the end.

\section{Complementarity relations}
%\section{Complementarity Relations}
\label{pq-relations} 
We now derive the complementarity relations for arbitrary multiparty quantum states. The parties involved are divided into ``legitimate'' users and ``eavesdroppers''. 
The relations show a trade-off between two quantities, one of which is the purity of the state of the legitimate users, while the other is a quantum characteristic 
in the legitimate users versus eavesdroppers bipartition. To be specific, we begin with the three party case, where there are two legitimate users and a single 
eavesdropper. We will briefly mention the case of an arbitrary number of parties later.

Consider therefore a three-party quantum system in the state $\rho_{ABC}$. 
For several non-classical bipartite correlation measures, \({\cal Q}'\), the relation ${\cal Q}'_{AB:C}-S_{AB} \leq 0$ holds, 
%\begin{equation}
%{\cal Q}'_{AB:C}-S_{AB} \leq 0,
%\end{equation} 
where $S_{AB}\equiv S(\rho_{AB})$ is the von Neumann entropy of its argument, and \(\rho_{AB} = \mbox{tr}_C \rho_{ABC}\). [\(S_{AC}\), etc. are similarly defined.] 
And \({\cal Q}'_{AB:C}\) is the non-classical correlation \({\cal Q}'\) of the state \(\rho_{ABC}\) in the \(AB:C\) bipartition.
The relation holds \cite{akash-meghla} e.g. for  entanglement of formation \cite{eof},
entanglement cost \cite{ent-cost}, distillable entanglement \cite{eof}, relative entropy of entanglement \cite{relent}, and one-way distilled key rate \cite{dw-rate}. 
Interestingly, \({\cal Q}'_{AB:C}\) can also be a measure of classical  correlation, as quantified by the measured quantum mutual information, defined as 
follows. 
The measured quantum mutual information \cite{hvoz}, of a bipartite quantum state \(\varrho_{XY}\) is defined as 
\(\mathcal{J}(\varrho_{XY}) = S(\rho_X) - \min \sum p_kS(\varrho_{XY}^k)\), where the minimization is over measurements performed by the party \(Y\) 
that creates 
the ensemble \(\{p_k, \varrho_{XY}^k\}\). Here, \(\rho_X = \mbox{tr}_Y \varrho_{XY}\). 
% The measured quantum mutual information is bounded above by the (unmeasured) quantum mutual information \cite{hvoz}, and 
% so the complementarity relations derived above are also valid for the measured quantum mutual information replacing the quantum mutual information.
That ${\cal Q}'_{AB:C}-S_{AB} \leq 0$ is valid for \({\cal Q}'\) identified with this classical correlation, follows e.g. from Ref. \cite{koashi-winter}.

\begin{widetext}
\begin{center}
\begin{figure}[htb]
{\includegraphics[width=2.1in, angle=0]{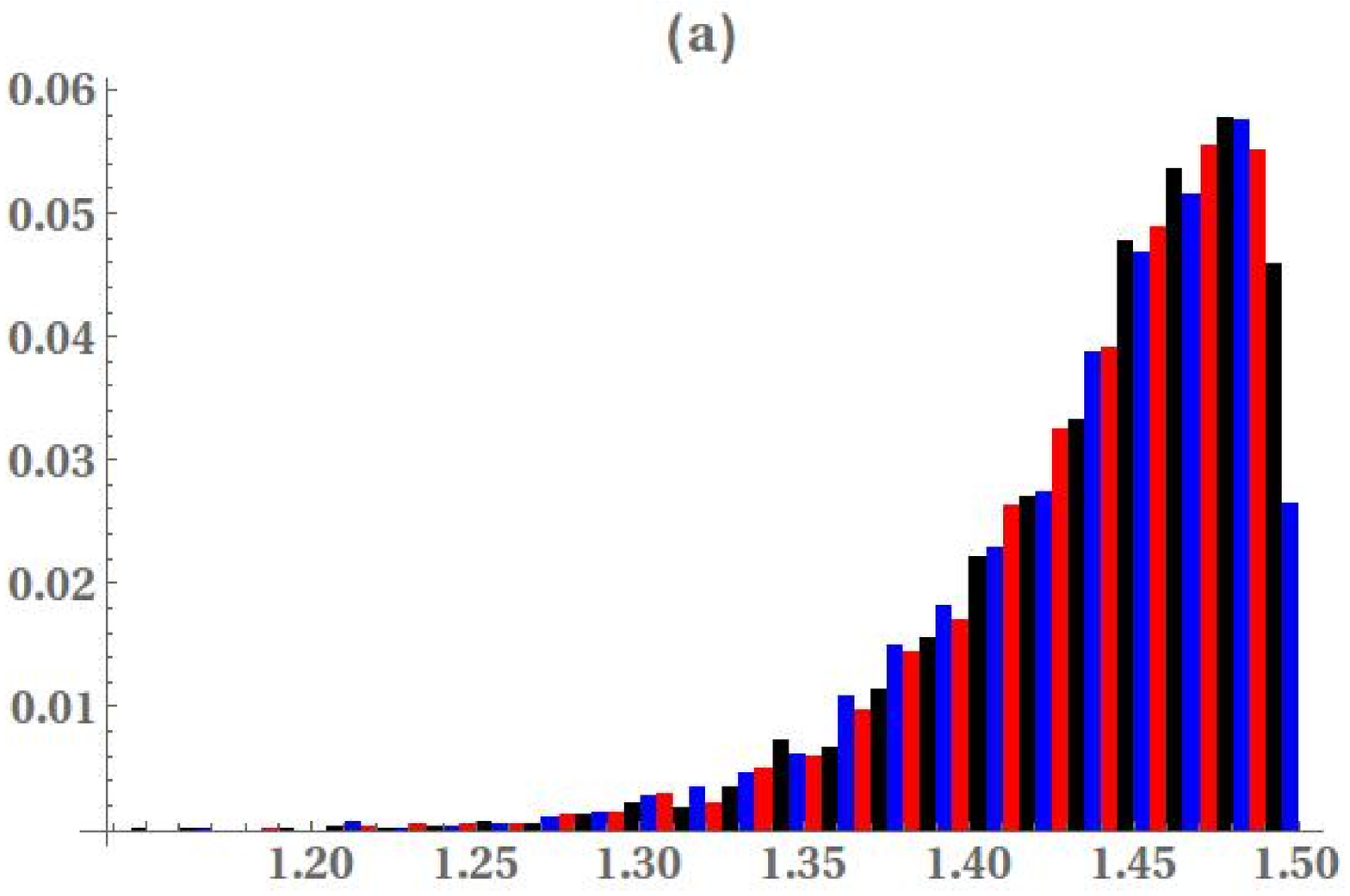}}
{\includegraphics[width=2.1in, angle=0]{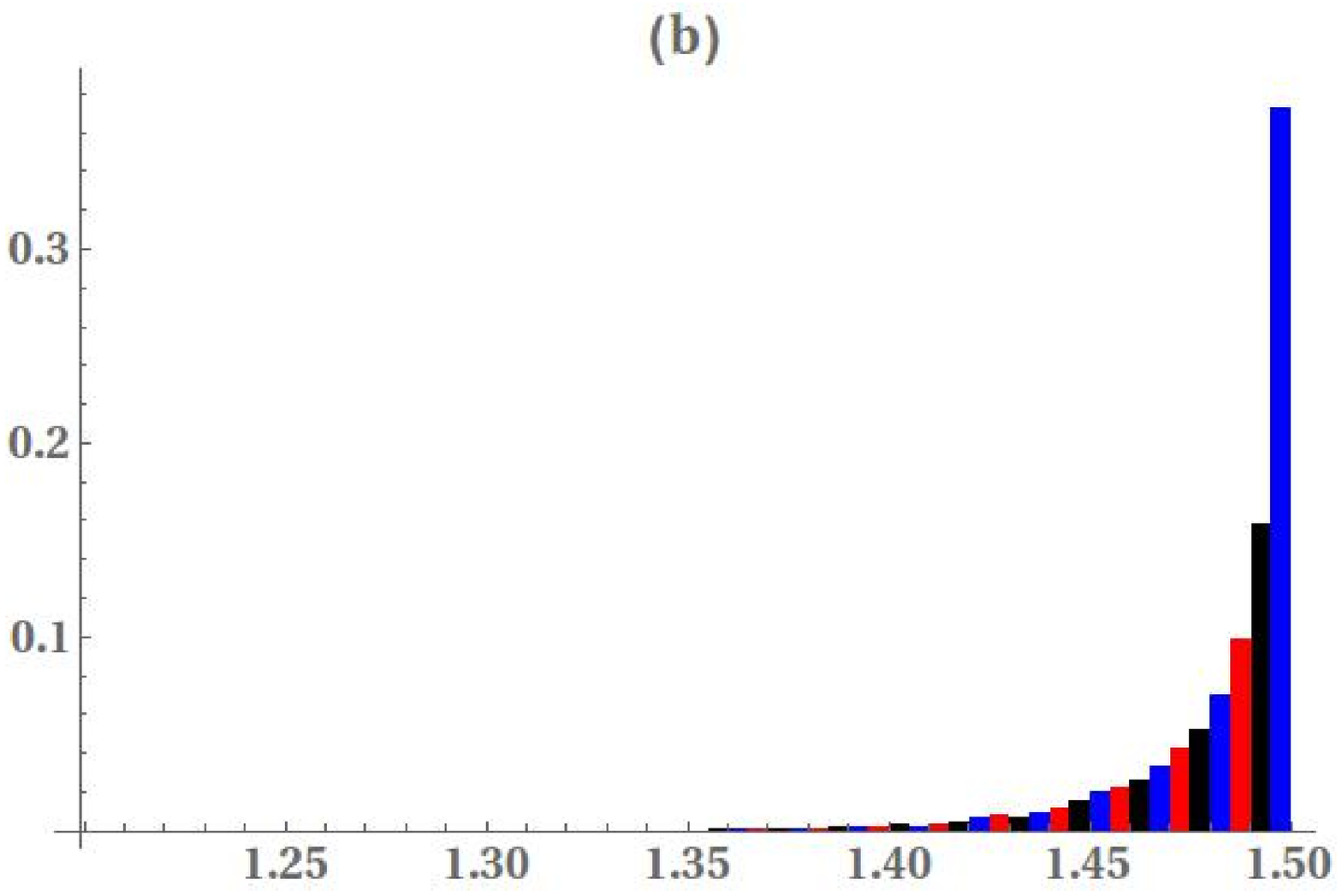}}
{\includegraphics[width=2.1in, angle=0]{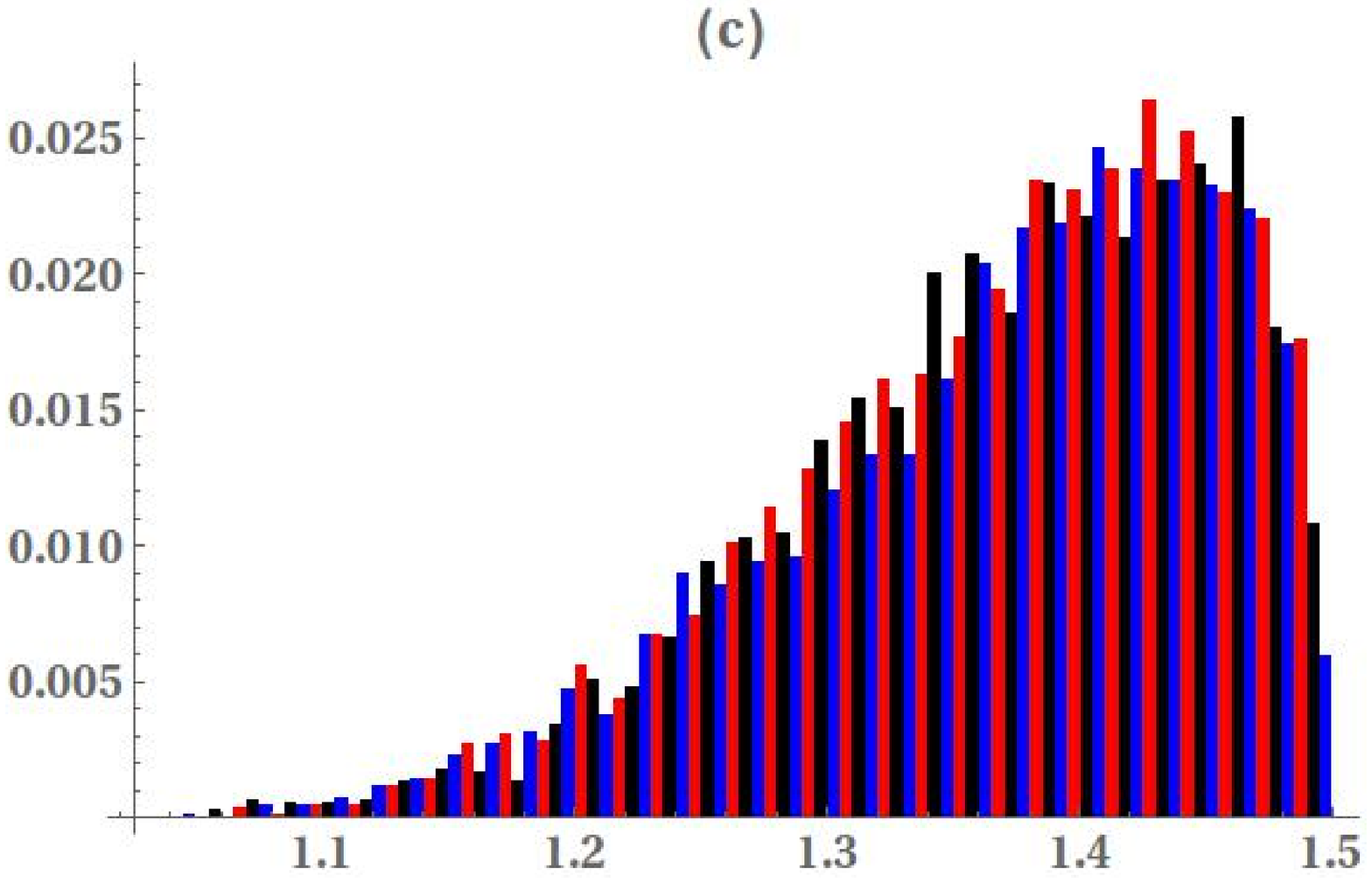}}
%{\includegraphics[width=1.6in, angle=0]{rank4-ent}}
\caption{(Color.)
The complementarity for three qubit systems. The different panels exhibit histograms for the sum of two quantities, viz. 
the normalized purity and a normalized correlation. The correlation 
is the negativity in panel (a), 
 logarithmic negativity in (b), and quantum mutual information in (c), for rank 1 states. We Haar uniformly generate \(10^4\) three qubit states for 
generating the histograms. 
The histograms for normalized measured quantum mutual information, quantum discord, and quantum work deficit are almost 
identical to that of the normalized quantum mutual information. The vertical axis represents the relative frequency of occurence of a randomly generated 
three-qubit state in the corresponding range of the sum of the two quantities on the horizontal axis. All quantities are dimensionless.
}
\label{fig:rank12345-ent}
\end{figure}
\end{center}
\end{widetext}

This relation ${\cal Q}'_{AB:C}-S_{AB} \leq 0$ can be rewritten, using appropriate normalized quantities, as
\begin{eqnarray}
\frac{\log_2d_{AB}-S_{AB}}{\log_2d_{AB}} + \frac{{\cal Q}'_{AB:C}}{\min\{\log_2d_{AB},\log_2d_C\}} \nonumber \\ \leq 1
+{\cal Q}'_{AB:C}\left(\frac{1}{\min\{\log_2d_{AB},\log_2d_C\}}-\frac{1}{\log_2d_{AB}}\right). \nonumber \\
\label{eq:qcr-def}
\end{eqnarray}
%where \(k\) is a positive number. 
While the above relation only needs ${\cal Q}'_{AB:C}-S_{AB} \leq 0$, we may note that the choice of the denominators in the terms on the left hand side have been 
guided by the fact that \(0\leq S(\rho_{AB}) \leq \log_2 d_{AB}\),  so that \(0\leq \log_2 d_{AB} - S(\rho_{AB}) \leq \log_2 d_{AB}\), and 
the oft-true relation  
\(0\leq {\cal Q}'_{AB:C} \leq \min\{\log_2 d_{AB},\log_2 d_C\} \).
We indicate the dimension of the Hilbert space corresponding to  a system denoted as \(X\) by \(d_X\). 

The first term, ${\cal P}_{AB}\equiv \frac{\log_2d_{AB}-S_{AB}}{\log_2d_{AB}}$, on the left hand side of ineq. (\ref{eq:qcr-def}) quantifies the purity of the system 
in the \(AB\) part, i.e. of \(\rho_{AB}\). We have normalized the quantity so that it varies between 0 and 1. 
The second term, ${\cal Q}_{AB:C}\equiv \frac{{\cal Q}'_{AB:C}}{\min\{\log_2d_{AB},\log_2d_C\}}$, represents the normalized non-classical correlation of the system 
in the \(AB:C\) bipartition. Again, it has been normalized, and if we assume that \(0\leq {\cal Q}'_{AB:C} \leq \min\{\log_2 d_{AB},\log_2 d_C\} \) is true, we once more 
have \( 0 \leq {\cal Q}_{AB:C} \leq 1\).
Thus the trivial upper bound of the quantity ${\cal P}_{AB}+{\cal Q}_{AB:C}$
%, the left hand side of (\ref{eq:qcr-def}), 
is 2. 
%We show below that the dimensions of subsystems modify and tighten the bound for different ${\cal Q}'$s.

%{\it For Quantum Correlation Measures}.--
Eq. (\ref{eq:qcr-def}) can further be shuffled into
%${\cal P}_{AB}+{\cal Q}_{AB:C} \leq 1$ when $d_{AB} \leq d_C$, and to ${\cal P}_{AB}+{\cal Q}_{AB:C} \leq 2-\frac{\log_2d_C}{\log_2d_{AB}}$ when ${\cal Q}'_{AB:C} \leq \log_2d_C$ in the case of $d_{AB} > d_C$.
%\[
% {\cal P}_{AB}+{\cal Q}_{AB:C} \leq
%  \begin{cases}
%   1 & \text{when } d_{AB} \leq d_C \\
%   2-\frac{\log_2d_C}{\log_2d_{AB}}       & \text{when } d_{AB} > d_C \\
%   & \text{and } {\cal Q}'_{AB:C} \leq \log_2d_C   
%  \end{cases}
%\]
\begin{align}
\label{eq:qcr-ent1}
{\cal P}_{AB}+{\cal Q}_{AB:C} \leq 1, & \text{ when } d_{AB} \leq d_C,
\end{align}
while it reads
\begin{align}
\label{eq:qcr-ent2}
{\cal P}_{AB}+{\cal Q}_{AB:C}& \leq 2-\frac{\log_2d_C}{\log_2d_{AB}}, \text{ when } d_{AB} > d_C \\
& \hspace{2.5cm} \text{ and } {\cal Q}'_{AB:C} \leq \log_2d_C. \nonumber
\end{align} 
% For entanglement measures like concurrence \cite{concurrence} and entanglement of formation (EoF) \cite{eof}, 
% the latter complementarity relation is saturated by $\frac{1}{\sqrt{2}}(|0\rangle^{\otimes n}_{AB}|0\rangle_C+|1\rangle^{\otimes n}_{AB}|1\rangle_C)$, 
% and the bound becomes two in the limit $n\rightarrow \infty$.
For systems in \(\mathbb{C}^d \otimes \mathbb{C}^d \otimes \mathbb{C}^d\), the complementarity relation reads 
\begin{equation}
\label{kuasha-ta}
{\cal P}_{AB}+{\cal Q}_{AB:C} \leq \frac{3}{2}.
 \end{equation}
Note that the right-hand-side is independent of dimension in this case.
In particular, for three-qubit quantum states, the bound is three-halves,
 and is saturated by the Greenberger-Horne-Zeilinger state \((|000\rangle + |111\rangle)/\sqrt{2}\) \cite{GHZ}. 
%
%Notice that for \(\mathbb{C}^2 \otimes \mathbb{C}^2 \otimes \mathbb{C}^d\) systems, the complementarity  reads 
%\begin{equation}
%{\cal P}_{AB}+{\cal Q}_{AB:C} \leq 2 - \frac{1}{2}\log_2 3.
% \end{equation}
% several classes of orthonormal Bell-like states \cite{asu-bell-like}, like
% \begin{align}
% |\psi\rangle^{1}_{x_1x_2x_3}&= \frac{1}{\sqrt{2}}(|0x_2x_3\rangle+(-1)^{x_1}|1\bar{x}_2\bar{x}_3), \nonumber \\
% |\psi\rangle^{2}_{x_1x_2x_3}&= \frac{1}{2}H^{\otimes 2}|x_1x_2\rangle|x_c \oplus x_3\rangle , \nonumber \\
% |\psi\rangle^{3}_{x_1x_2x_3}&= \frac{(-1)^{x_1x_2}}{2}H^{\otimes 2}|x_1x_2\rangle|x_c \oplus x_3\rangle ,
% \label{eq:asu-bell-like} 
% \end{align}
% where $x_i \in \{0,1\}$, $\bar{x}_i=x_i \oplus 1$, 
%  $ H=\frac{1}{\sqrt{2}}\left(
%   \begin{array}{cc}
%      1 & 1\\
%      1 & -1
%   \end{array}
%   \right)$ is single-qubit Hadamard operator, and $x_c=x_1\oplus x_2$.\\  
As mentioned before, the complementarity holds for several entanglement measures. 
It is natural to ask whether the same holds for other quantum correlation measures 
%We are willing to conjecture that similar relations hold for information-theoretic quantum correlation measures
including entanglement measures like negativity \cite{lneg} and logarithmic negativity \cite{lneg}, 
and information theoretic quantum correlations like quantum discord \cite{hvoz} and quantum work-deficit \cite{work-deficit}.
If quantum discord is defined by considering the measurement in its definition to be in the first party, then its value for an arbitrary quantum state in 
\(AB:C\) can be shown to be bounded above by \(S_{AB}\) \cite{hari-he-madhabo-chulkabo-na-gha-dobo}, so that 
the complementarity relation  (\ref{eq:qcr-ent1}) is valid for the measure. Henceforth, we consider that the quantum discord is defined by performing the measurement 
in the second party. 
Numerical simulations by Haar uniform generation of three-qubit states having different ranks (ranging from 1 till 4) show that several correlation measures
%information-theoretic quantum correlation measures 
indeed obey the above complementarity (ineq. (\ref{kuasha-ta})). See Figs. \ref{fig:rank12345-ent} and \ref{fig:rank12345-qc} for 
depictions of the cases of ranks 1 and 2. For higher ranks, the complementarities become less and less tight. 
We will discuss in the succeeding section about the tightness of these 
relations. 
It is also clear that similar complementarities will hold also when purity is quantified by utilizing R{\'e}nyi \cite{Renyi} and Tsallis \cite{Tsallis} 
entropies.

\begin{widetext}
\begin{center}
\begin{figure}[htb]
{\includegraphics[width=2.1in, angle=0]{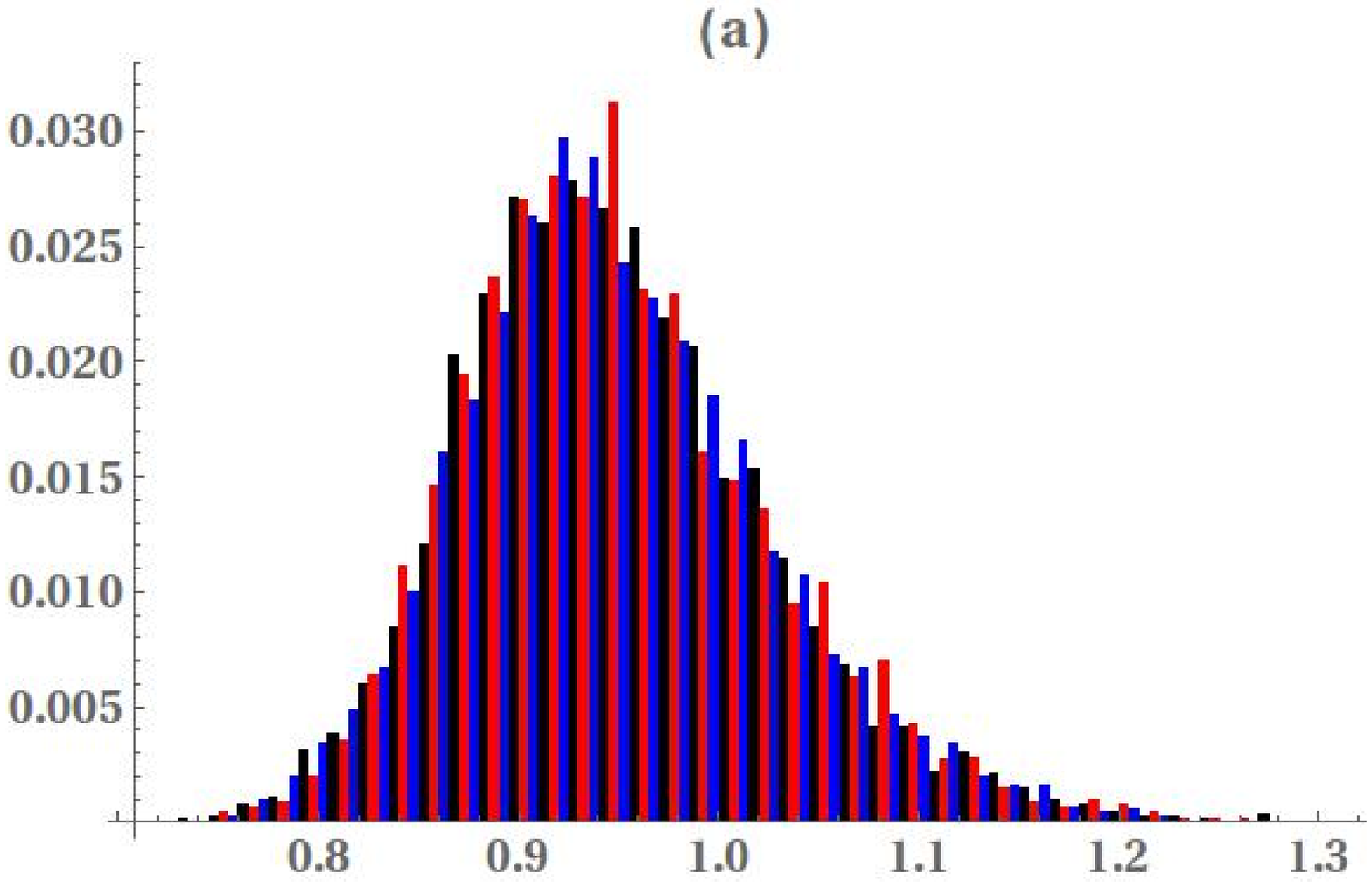}}
{\includegraphics[width=2.1in, angle=0]{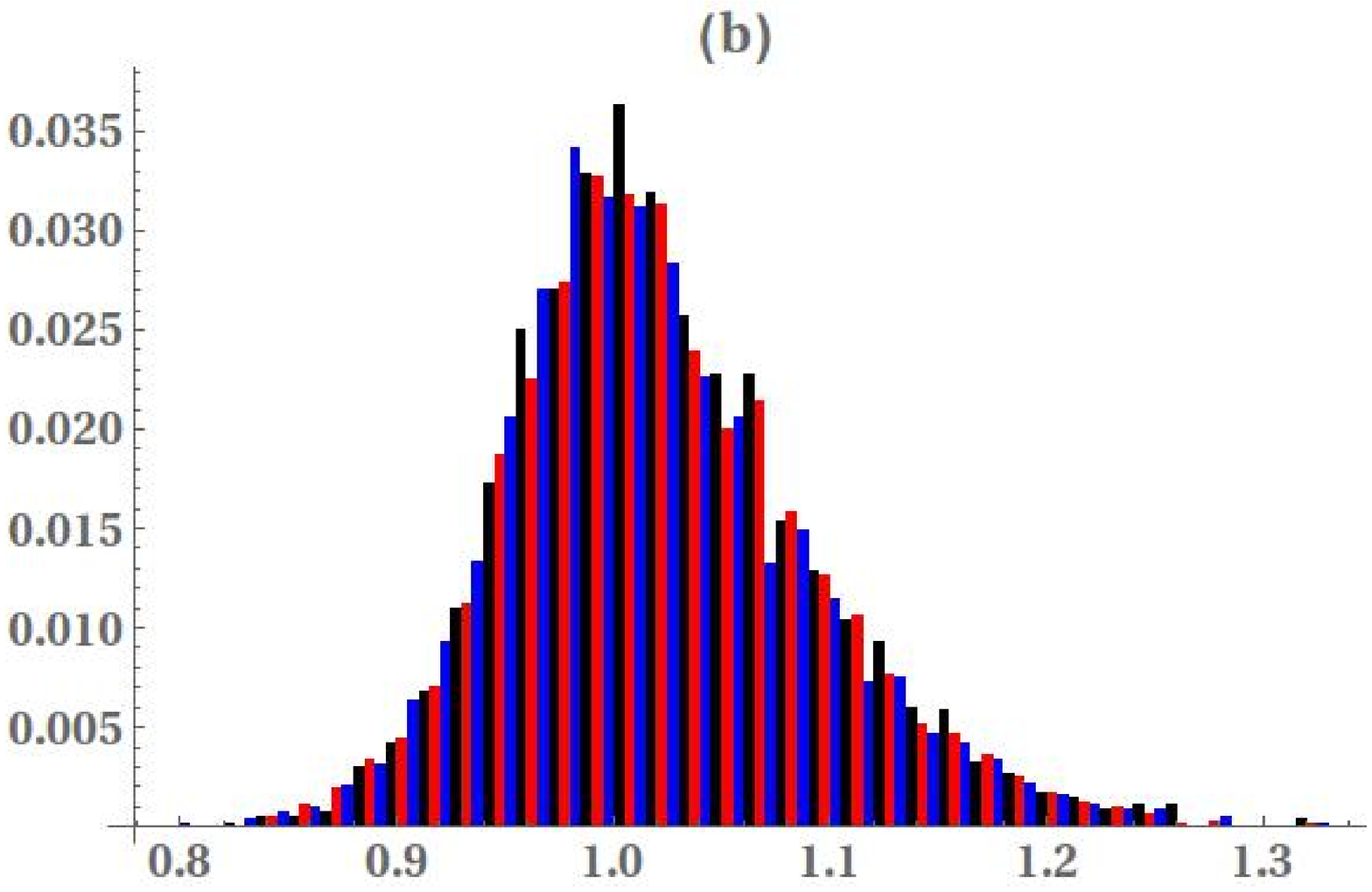}}
{\includegraphics[width=2.1in, angle=0]{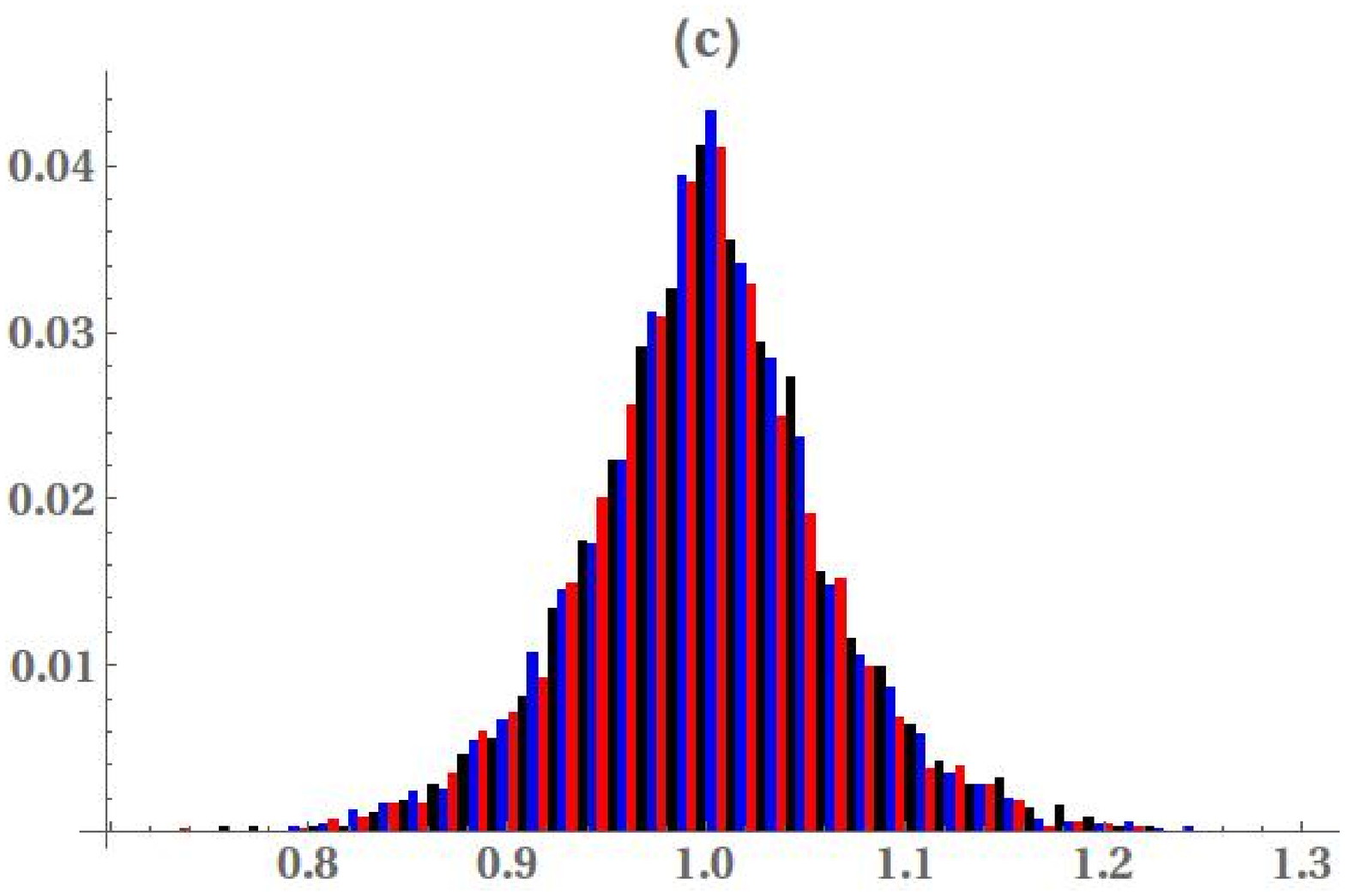}}\\
{\includegraphics[width=2.1in, angle=0]{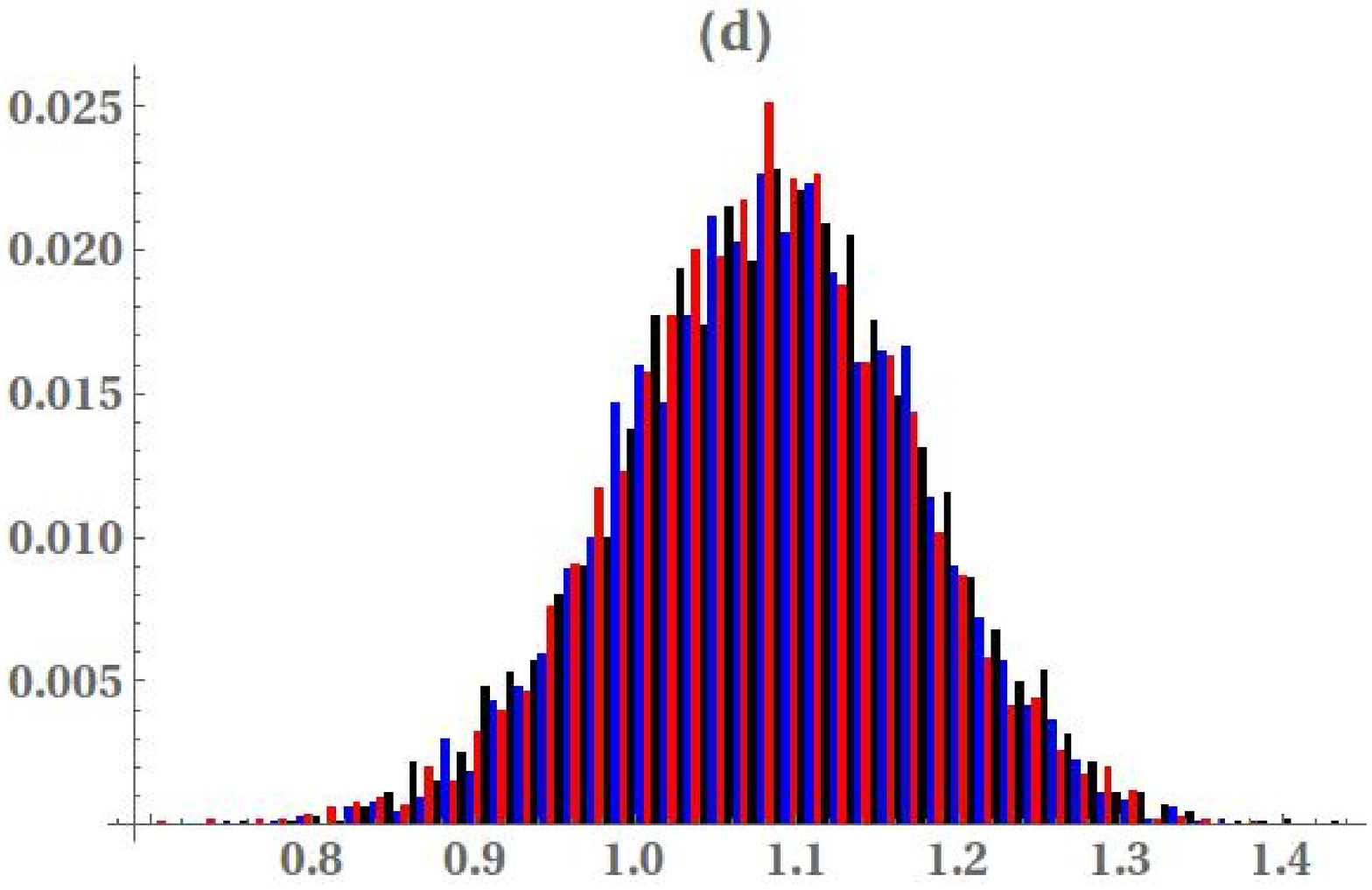}}
{\includegraphics[width=2.1in, angle=0]{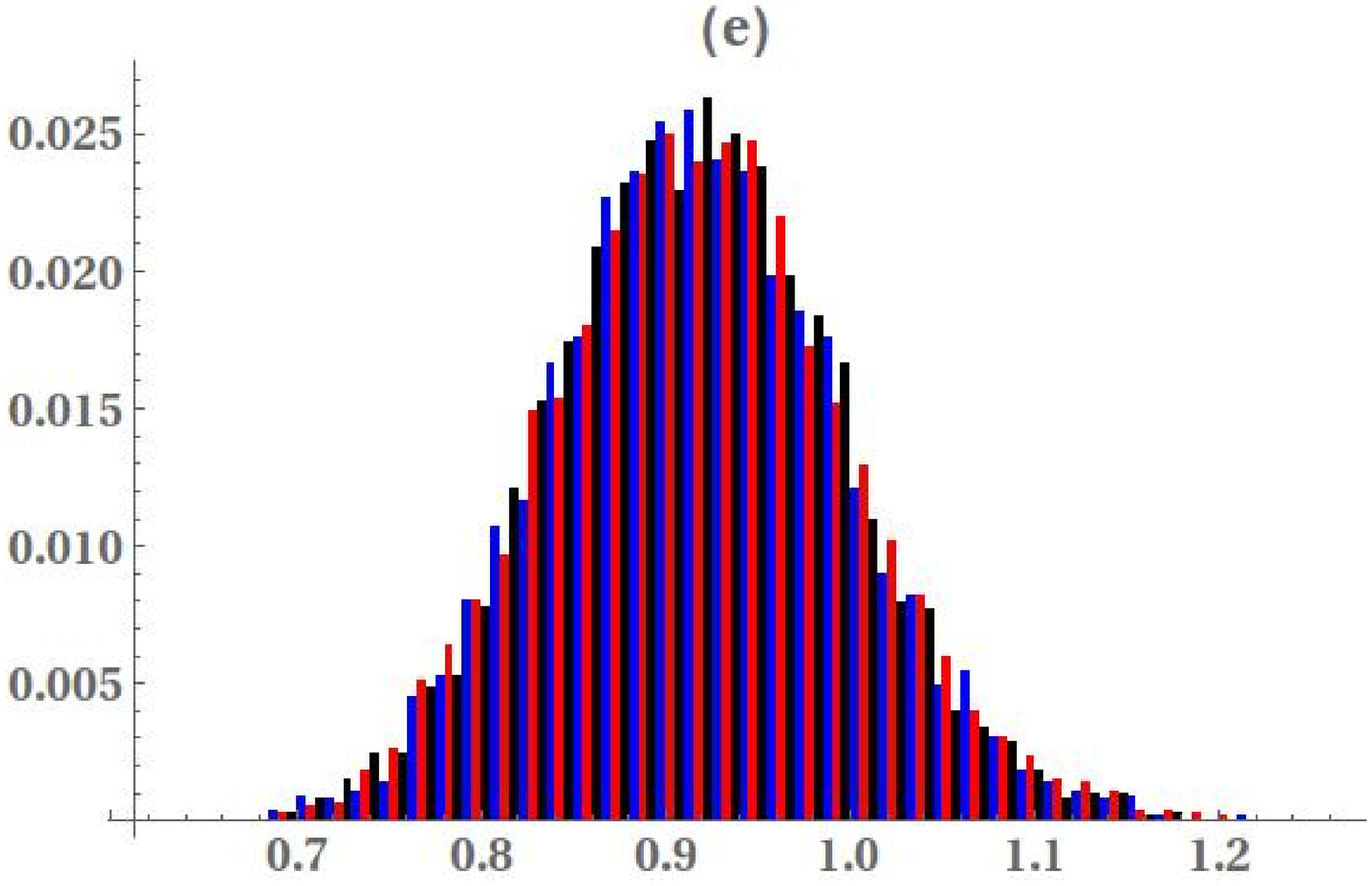}}
{\includegraphics[width=2.1in, angle=0]{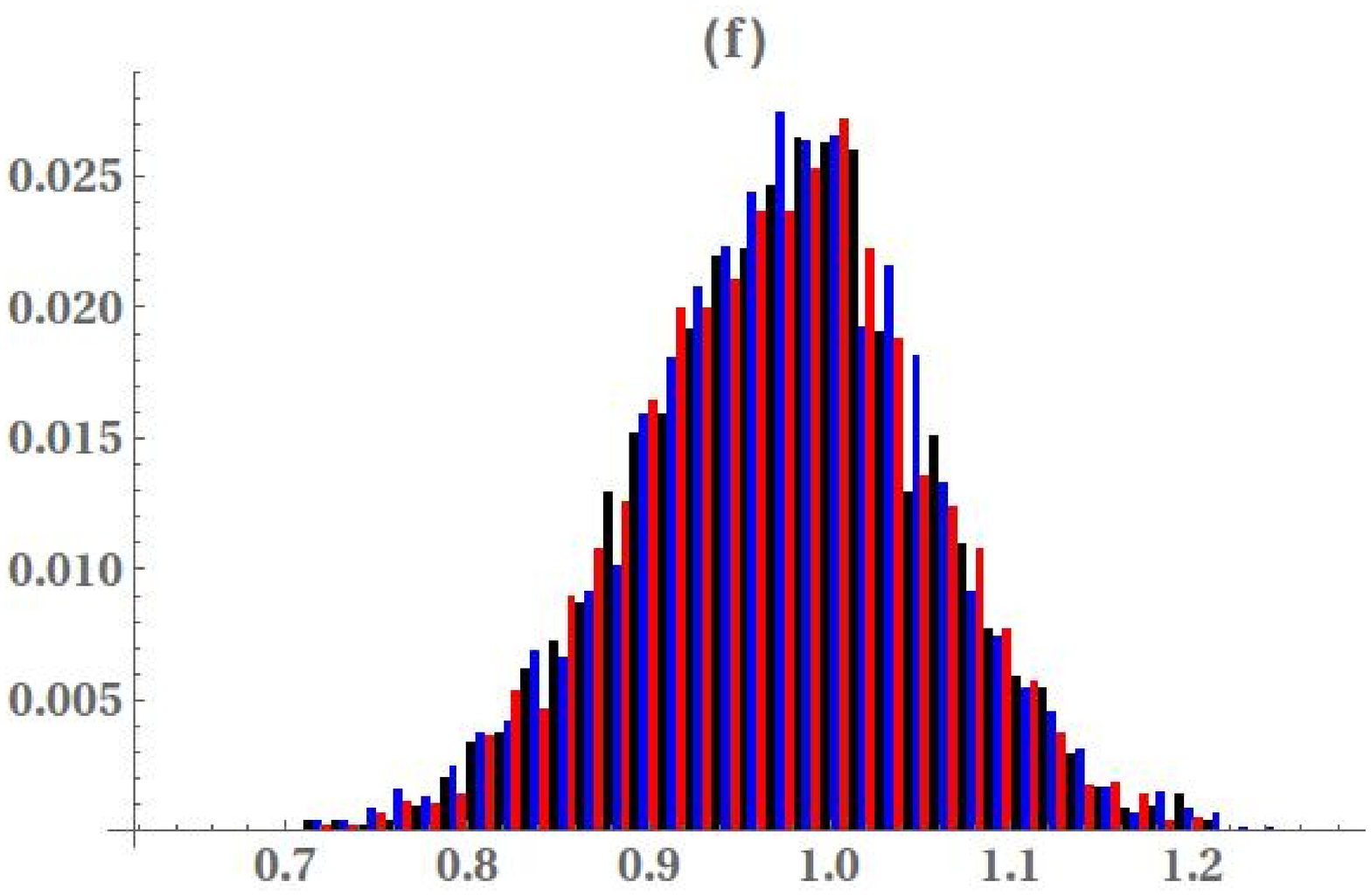}}
\caption{(Color.)
The complementarity relation, again for three qubit systems, but for rank 2 states. 
The different panels exhibit histograms for the sum of two quantities, viz. 
the normalized purity and a normalized correlation. The correlation 
is the negativity in panel (a), 
 logarithmic negativity in (b), quantum mutual information in (c), measured quantum mutual information in (d), 
quantum discord in (e), and quantum work deficit in (f). 
All other considerations are the same as in Fig. \ref{fig:rank12345-ent}.
}
\label{fig:rank12345-qc}
\end{figure}
\end{center}
\end{widetext}

%\subsection{For Mutual Information}
The quantum mutual information  \cite{thomas-cover, qmi-bunch} of a bipartite quantum state $\varrho_{XY}$, identified with the 
total correlation in the state, is defined as $I'_{X:Y}=S_X+S_Y-S_{XY}$.
%\begin{align}
%I'(X:Y)&=S_X+S_Y-S_{XY}.
%\end{align}
Using the Araki-Lieb triangle inequality \cite{Araki-Lieb}, $|S_X-S_Y| \leq S_{XY}$, it follows that 
%$I'(X:Y) \leq 2\min\{S_X,S_Y\} \leq 2\min\{\log_2d_X,\log_2d_Y\}$.
\begin{align}
I'_{X:Y}& \leq 2\min\{S_X,S_Y\} \leq 2\min\{\log_2d_X,\log_2d_Y\}.
\label{eq:qmi-bound}
\end{align}
The Araki-Lieb inequality therefore helps us to obtain the relations in (\ref{eq:qcr-ent1}) and (\ref{eq:qcr-ent2}) with the normalized 
%non-classical 
correlation \({\cal Q}\)
replaced by the normalized  quantum mutual information, with the additional property that the parallel of the condition \({\cal Q}'_{AB:C} \leq \log_2d_C\) 
being automatically satisfied here. More precisely, we have 
\begin{align}
\label{eq:qcr-qmi}
{\cal P}_{AB}+{\cal I}_{AB:C} \leq 1  \text{ when } d_{AB} \leq d_C, \\
{\cal P}_{AB}+{\cal I}_{AB:C} \leq 2-\frac{\log_2d_C}{\log_2d_{AB}} \text{ when } d_{AB} > d_C,
\label{eq:qcr-qmi-abar}
\end{align}
where 
\({\cal I}_{AB:C} = \frac{I'_{AB:C}}{2\min\{\log_2d_{AB},\log_2d_C\}}\).

% When $d_{AB} \leq d_C$, the left hand side of Eq. (\ref{eq:qcr-def}) becomes  
% \begin{align}
% {\cal P}_{AB}+{\cal I}_{AB:C}&= \frac{2\log_2d_{AB}-(S_{AB}+S_{(AB)C}-S_C)}{2\log_2d_{AB}} \nonumber \\
% & \leq 1,
% %\label{eq:qcr-qmi1}
% \end{align}
% using Lieb's inequality. This bound is trivially saturated by pure states.
% On the other hand when $d_{AB} > d_C$, since ${\cal I}'_{AB:C}\leq 2\log_2d_C$, we have   
% \begin{align}
% {\cal P}_{AB}+{\cal I}_{AB:C}& \leq 2- \frac{(S_{AB}+S_{(AB)C}-S_C)+2\log_2d_{C}}{2\log_2d_{AB}} \nonumber \\
% & \leq 2- \frac{\log_2d_{C}}{\log_2d_{AB}},
% \label{eq:qcr-qmi2}
% \end{align}
% using Lieb's inequality. 
For quantum states of three qudits, we again have the dimension-independent complementarity bound,
\begin{equation}
{\cal P}_{AB}+{\cal I}_{AB:C} \leq \frac{3}{2},
 \end{equation}
and again for three qubits, the GHZ state saturates the bound.
%This bound is again saturated by $\frac{1}{\sqrt{2}}(|0\rangle^{\otimes n}_{AB}|0\rangle_C+|1\rangle^{\otimes n}_{AB}|1\rangle_C)$.
%and in the limit $n\rightarrow \infty$ the bound becomes two.
%For three-qubit quantum states, it is three-half and is achieved by classes of states in Eq. (\ref{eq:asu-bell-like}).

The (classical) mutual information \cite{thomas-cover} between two observables \({\cal X}\) and \({\cal Y}\) of the systems \(X\) and \(Y\) is given by 
\begin{equation}
\tilde{I}'_{{\cal X}: {\cal Y}} = H({\cal X}) + H({\cal Y}) - H({\cal X},{\cal Y}), 
\end{equation}
where \(H({\cal X}) = -\sum_i p^{\cal X}_i \log_2 p^{\cal X}_i\) is the Shannon entropy of the observable \({\cal X}\), 
with the observable \({\cal X}\) having been measured in the state \(\varrho_{XY}\) and the outcomes 
\(x_i\) obtained with the Born probabilities \(p^{\cal X}_i = \mbox{tr}_X(\mbox{tr}_Y(\varrho_{XY}){\cal X})\). \(H({\cal Y})\) is similarly defined, and 
\(H({\cal X},{\cal Y})\) is the joint entropy of the observable \({\cal X} \otimes {\cal Y}\) when measured in the state \(\rho_{XY}\). 
Now the classical mutual information is bounded above by the quantum one \cite{molmer}, and so the relations for the quantum mutual information derived above 
are also true for the classical variety. We will use the notation 
\(\tilde{I}_{{\cal X}: {\cal Y}} = \frac{\tilde{I}'_{{\cal X}: {\cal Y}}}{2\min\{\log_2 d_{X}, \log_2 d_Y\}}\).

% The measured quantum mutual information \cite{hvoz}, of a bipartite quantum state \(\rho_{AB}\) is defined as 
% \(\mathcal{J}(\rho_{AB}) = S(\rho_B) - \min \sum p_kS(\rho_{AB}^k)\), where the minimization is over measurements performed by the party \(A\) that creates 
% the ensemble \(\{p_k, \rho_{AB}^k\}\). Here, \(\rho_B = \mbox{tr}_A \rho_{AB}\). The measured quantum mutual information is bounded above by the (unmeasured) quantum mutual information \cite{hvoz}, and 
% so the complementarity relations derived above are also valid for the measured quantum mutual information replacing the quantum mutual information.

Since quantum mutual information is non-increasing under discarding of parties \cite{qmi-bunch}, 
and since classical mutual information is a lower bound for the quantum one \cite{molmer},
the quantum mutual information in the \(AB:C\) bipartition, in the complementarity, can be replaced by the minimum of the quantum or classical mutual information 
in \(A:C\) and \(B:C\). Similarly, the quantum correlation in the \(AB:C\) partition can be replaced by the minimum of the quantum correlation in \(A:C\) and \(B:C\), 
by assuming that the quantum correlation is non-increasing under discarding of parties. 
% We will come back to these relations when we relate a complementarity with 
% the security proof in a quantum cryptographic protocol. Moreover, in the next section, we will provide numerical evidence that the complementarity between purity 
% and \(\min\{\mathcal{Q}_{A:C},\mathcal{Q}_{B:C}\}\) can be bettered. 

%[COMMENT: WHAT WAS THE RELATION WITH DISCORD HERE???]

% The above bounds can also be obtained from inequalities (\ref{eq:qcr-ent1}) and (\ref{eq:qcr-ent2}) for ${\cal Q}'_{AB:C}=\frac{1}{2}{\cal I}'_{AB:C}$ 
% since $\frac{1}{2}{\cal I}'_{AB:C} \leq S_{AB}$ and $\frac{1}{2}{\cal I}'_{AB:C} \leq \log_2d_C$ from Eq. (\ref{eq:qmi-bound}).\\
% From the non-negativity of quantum discord \cite{hvoz}, we know that classical mutual information (CMI) is bounded from above by QMI. Hence, the above 
% analysis also holds for classical mutual information. Thus we notice that low-dimensional eves can pilfer information efficiently, and hence can be fatal for 
% secure communication. 

As mentioned earlier, the complementarity relations are also true for \(N\)-party systems in the following sense. We envisage a quantum communication protocol
of \(N-r\) ``legitimate'' parties, and  a further \(r\) ``eavesdroppers'' who are trying to obtain some information from the legitimate users. 
The entire system of \(N\) parties share a quantum state \(\rho\). It is now possible to obtain complementarities in this \(N\)-party system between 
purity of the state of the \(N-r\) legitimate users and correlation in the legitimate users versus eavesdroppers bipartition.

It is clear that the complementarity relations considered will also be true for \(\mathcal{P} + \min\{\mathcal{Q}_{A:C},\mathcal{Q}_{B:C}\}\), 
provided \(\mathcal{Q}\) is non-increasing under discarding of subsystems. Numerical analysis of this quantity for 
\(10^4\) Haar uniformly generated states of three qubits, separately for rank-1, 2, and 3 states, reveals that it is indeed true for all the correlation measures 
considered. See Fig. \ref{fig:khoRer-chal}.
%\#\#\#\#\#\#
%  a stronger complementarity, viz. 
% \(\mathcal{P} + \min\{\mathcal{Q}_{A:C},\mathcal{Q}_{B:C}\}\) is upper bounded by unity. We have carried out the analysis separately for  rank-1 and rank-2 
% three-qubit states. We used the logarithmic negativity \cite{lneg} as the measure of quantum correlation.

\begin{widetext}
\begin{center}
\begin{figure}[htb]
\subfloat[rank-1]{\includegraphics[width=2in, angle=0]{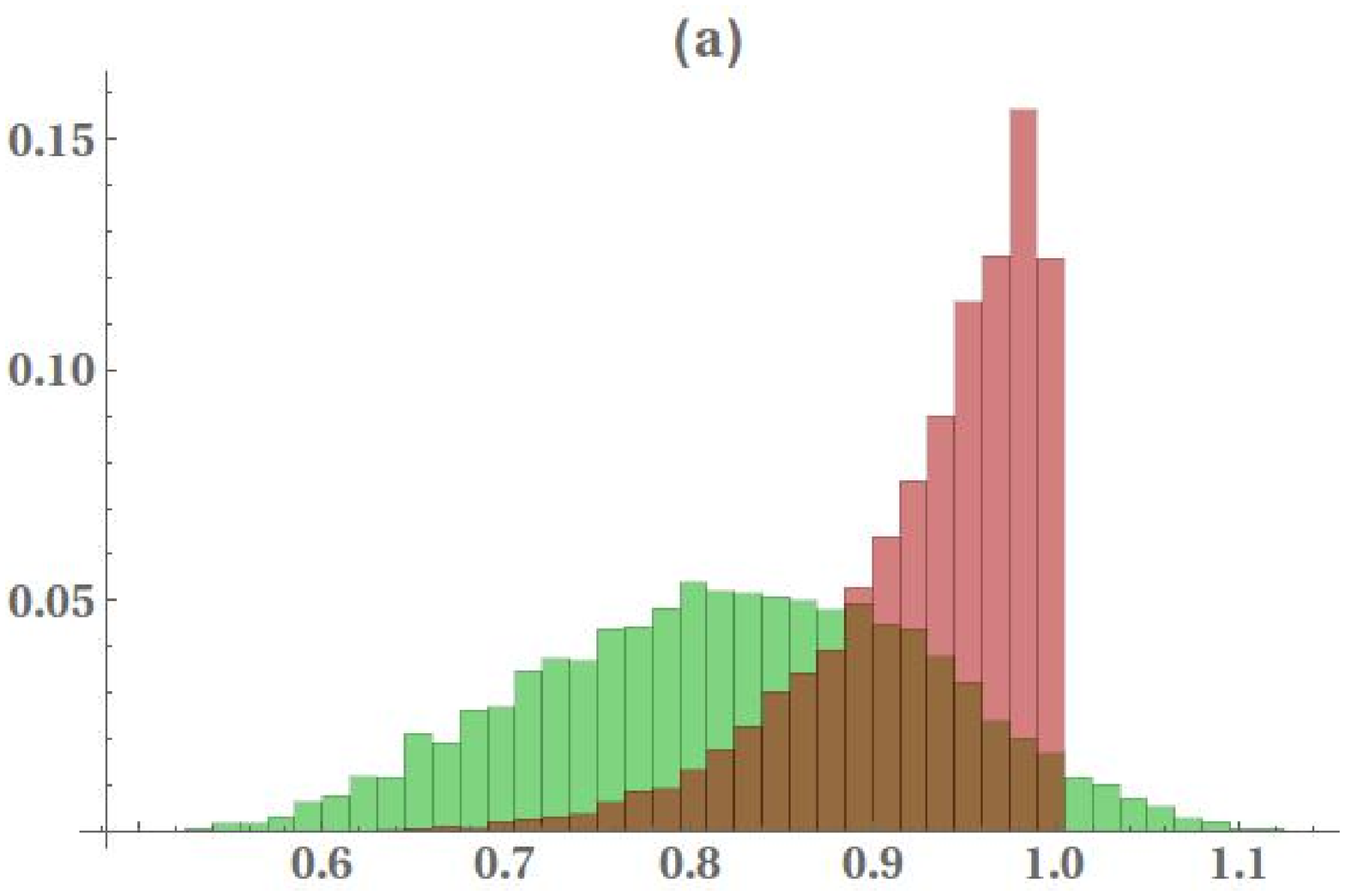}}
\subfloat[rank-2]{\includegraphics[width=2in, angle=0]{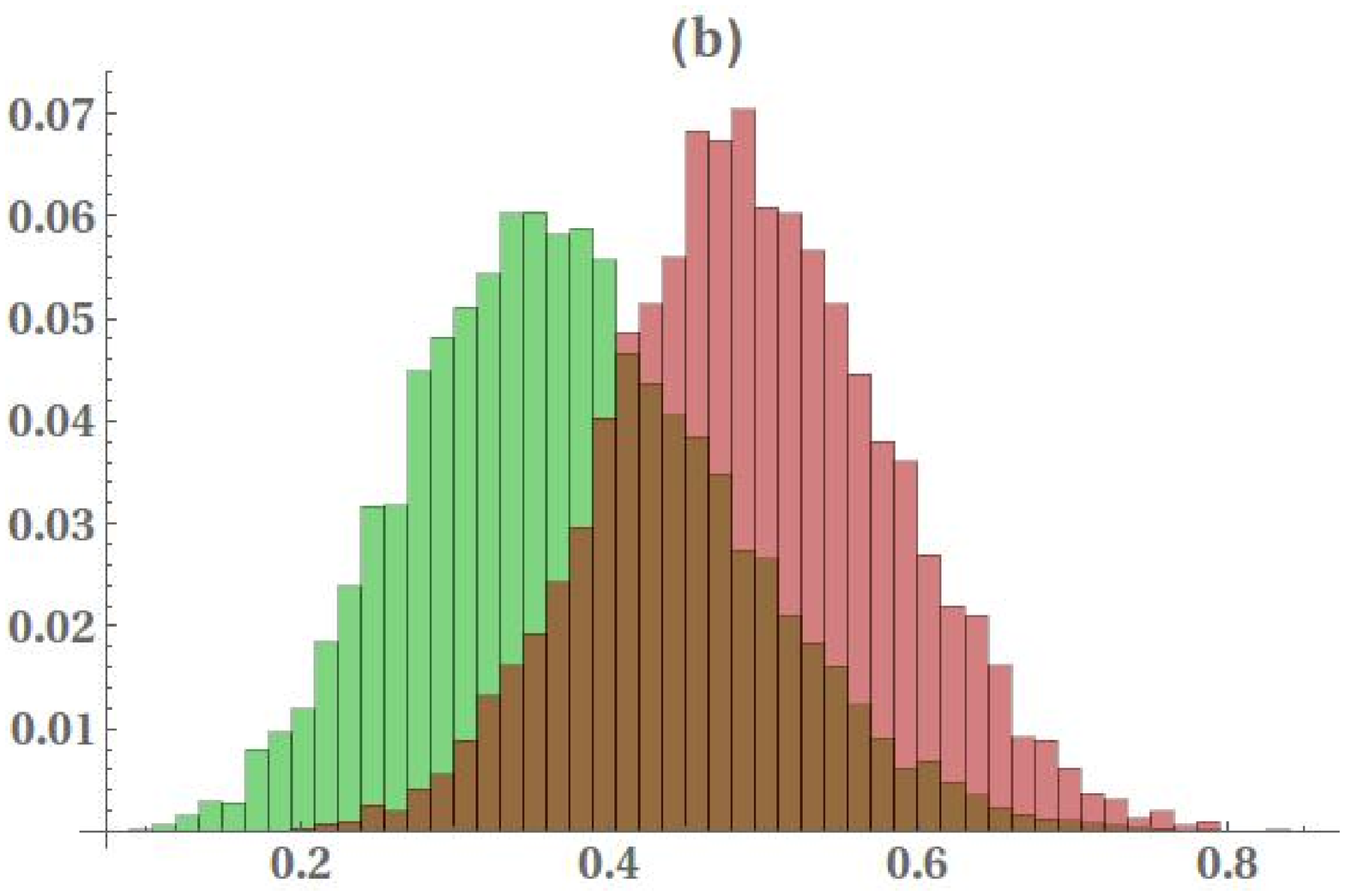}}
\subfloat[rank-3]{\includegraphics[width=2in, angle=0]{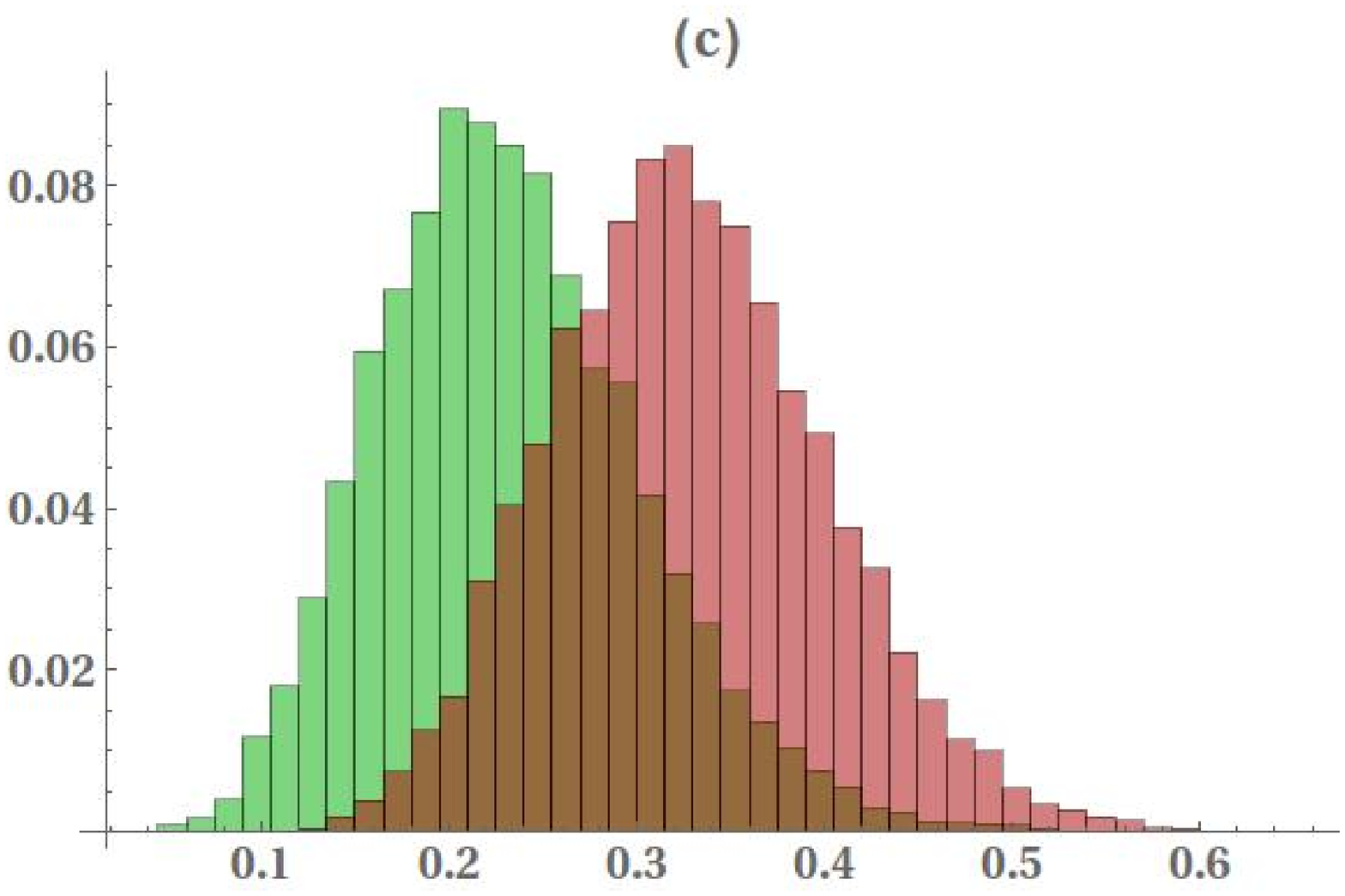}}
% \subfloat[rank-4]{\includegraphics[width=1.6in, angle=0]{rank4-ent}}
\caption{(Color.)
The quantity \(\mathcal{P} + \min\{\mathcal{Q}_{A:C},\mathcal{Q}_{B:C}\}\).
The different panels exhibit overlapping histograms, of three qubit states of different ranks, for this quantity.
% that violate this inequality. 
The green (red) bars correspond to the cases when \(\mathcal{Q}\) represents the normalized negativity (normalized quantum mutual information).
The panels (a), (b), and (c) are respectively for ranks 1, 2, and 3.
For each rank, we Haar uniformly generate $10^4$ three qubit states of that rank.
The vertical axis represents the relative frequency of occurence of a randomly generated 
three-qubit state of the considered rank in the corresponding range of the sum of the two quantities on the horizontal axis.
All quantities are dimensionless.
}
\label{fig:khoRer-chal}
\end{figure}
\end{center}
\end{widetext}

\section{Tightness of the complementarity relations}
%\section{Tightness of Complementarity Relations}
\label{tightness}
%The complementarity relations have been extablished for several 
%We have shown above that for $\mathbb{C}^{d_A}\otimes\mathbb{C}^{d_B}\otimes \mathbb{C}^{d_C}$ quantum systems, when $d_{AB}>D_C$, 
%the complementarity relation  ${\cal P}_{AB}+{\cal Q}_{AB:C} \leq 2-\frac{\log_2d_C}{\log_2d_{AB}}$ between normalized purity ${\cal P}_{AB}$ 
%and normalized quantum correlation ${\cal Q}_{AB:C}$, holds for several 
%nonclassical  measures.
%  like entanglement of formation \cite{eof}, 
% distillable entanglement \cite{eof}, quantum mutual information \cite{thomas-cover, qmi-bunch}, and measured quantum mutual information \cite{hvoz}. 
% We would
% %wish to 
% now like to investigate  
We now investigate how tight the complementarity relations are for the different measures.
%, and (b) 
%whether further quantum correlation measures like negativity \cite{lneg}, logarithmic negativity \cite{lneg}, quantum discord, 
%and quantum work-deficit satisfy them. 
%the same complementarity relation. 
%We Haar uniformly generate quantum  states of different ranks on 
%generate 
%$10^4$ random 
%$\mathbb{C}^{2}\otimes\mathbb{C}^{2}\otimes \mathbb{C}^{2}$, and create scatter diagrams of the measures versus purity (both normalized) for 
%these states. 
% $\rho_{ABC}^{(r)}$ 
%of varying ranks \(r\), using Haar uniform distribution, and 
%The plots are presented in the different panels in Figs. 
%The  negativity, logarithmic negativity, quantum mutual information, 
%measured mutual information, quantum discord and quantum work-deficit. The scatter diagrams of these (normalized) correlation measures against the (normalized)
 %purity are shown in Fig. 
%\ref{fig:rank12345-qc} and 
%\ref{fig:rank12345-ent} and \ref{fig:rank12345-qc}.
% and  \ref{fig:rank12345-qc}. 
%We observe that all quantum measures computed obey the corresponding complementarity relation. 
%${\cal P}_{AB}+{\cal Q}_{AB:C} \leq \frac32$. 
To examine the tightness of the  relation, we compute the average perpendicular distance of the points
%, in the scatter diagrams in the panels in Figs. 
%\ref{fig:rank12345-ent} and \ref{fig:rank12345-qc},
%(${\cal P}_0,{\cal Q}_0$) for each quantum state,
from the straight line representing the corresponding complementarity relation. 
The average is performed over the uniform Haar distribution for every considered rank, and over the 
corresponding quantum state space (for that rank).  
%${\cal P}+{\cal Q}=\frac32$, and take their average. The perpendicular distance of the point ($x_0,y_0$) from the straight line $ax+by+c=0$ is given 
%by $p=\frac{|ax_0+by_0+c|}{\sqrt{a^2+b^2}}$.
%\begin{equation}
%p=\frac{|ax_0+by_0+c|}{\sqrt{a^2+b^2}}.
%\end{equation}
In Table \ref{table:dist-222-r1-r5}, we list the average perpendicular distances for the different measures for states of different ranks. 
%  of point $({\cal P},{\cal Q})$ for the aforementioned quantum correlation measures 
% %like negativity, logarithmic negativity, QMI, MMI, quantum discord and quantum work-deficit,
% from the complementarity line ${\cal P}+{\cal Q}=\frac32$, of three-qubit quantum states $\rho_{ABC}^{(r)}$ of rank-\(r\), $r=1,2,3,4$. 
We see that 
rank-\(1\) (pure) quantum states  satisfy the complementarity relations quite tightly, while  it becomes comparatively weaker 
with the increase in the rank. See also Figs. 
\ref{fig:rank12345-ent} and \ref{fig:rank12345-qc} in this regard. Note that increasing the rank of the density
 matrix, typically,  increases its mixedness, driving it towards the (normalized) identity matrix, for which the sum of the purity and any of the correlation 
measures considered
vanishes. 
%is zero.  
%${\cal P}+{\cal Q}=0$. 
It is therefore expected that the average perpendicular distance would increase with 
increasing rank, 
as also 
%and this is what we also 
observed
 in Table \ref{table:dist-222-r1-r5}. 
%In Table \ref{table:dist-rank1}, in Appendix, we list the average perpendicular distance of point $({\cal P},{\cal Q})$ for quantum correlation measures like negativity, logarithmic negativity and QMI, from the complementarity line ${\cal P}+{\cal Q}=1$, of rank-\(1\) three-party quantum states $\rho_{ABC}\in {\cal H}^{2 \otimes 2 \otimes 3}$, $\rho_{ABC}\in {\cal H}^{2 \otimes 2 \otimes 4}$ and $\rho_{ABC}\in {\cal H}^{2 \otimes 2 \otimes 5}$. 

%%%
\begin{center}
\begin{table}[h]
\begin{tabular}{|c|c|c|c|c|}
\hline
         & rank-1 & rank-2 & rank-3 & rank-4 \\ \hline \hline
${\cal N}$ & 0.043   & 0.390     & 0.590  & 0.706     \\ \hline
$E_{\cal N}$ & 0.013    & 0.338   & 0.531  & 0.651   \\ \hline
${\cal I}$ & 0.093   & 0.354    & 0.509  & 0.612            \\ \hline
$\mathcal{J}$ & 0.093    & 0.296    & 0.481  & 0.605            \\ \hline
${\cal D}$ & 0.093   & 0.412    & 0.536  & 0.619     \\ \hline
$\vartriangle$ & 0.093   & 0.372    & 0.504  & 0.594 \\ \hline
\end{tabular}
\caption{Average perpendicular distance of the point $({\cal P},\tilde{{\cal Q}})$ from the line  ${\cal P}+\tilde{{\cal Q}}=\frac32$, where 
\(\tilde{{\cal Q}}\) represents the normalized versions of negativity (\(\mathcal{N}\)), or 
logarithmic negativity (\(E_\mathcal{N}\)), or quantum mutual information (${\cal I}$), or measured quantum mutual information ${\cal J}$, 
or quantum discord (\(\mathcal{D}\)), or 
quantum work-deficit (\(\vartriangle\)).
The different columns represent values for different ranks of the three qubit states, while the 
different rows are for different measures. For every rank, \(10^4\) states are generated Haar uniformly.
}
\label{table:dist-222-r1-r5}
\end{table}
\end{center}
%%%

\section{Application in quantum cryptography}
\label{sec-crypto}

Let us now discuss whether the derived complementarity relations can have implications in quantum information protocols. 

{\it The QKD setup}.-- We consider the QKD protocol proposed in Ref. \cite{acin-protocol} (see also \cite{acin-diqkd}) 
that is a modification of the Ekert 1991 protocol \cite{ekert91}. Suppose that Alice and Bob share a two-party state, and  
%quantum channel consisting of a source that emits pairs of entangled particles. 
Alice chooses between the measurement settings $A_0$, $A_1$, and $A_2$, while Bob chooses between $B_1$ and $B_2$ on their respective portions of the shared state.
% as possible measurements on each of their particles. 
Each  measurement is assumed to have two  outcomes. 
%The actual dimensions of the systems in possession of Alice and Bob may be higher. 
The measurement results of \(A_0\) and \(B_1\) are used to obtain the raw key, and 
% % labeled 
% %by $a_i,b_j \in  \{\pm 1\}$ (the quantum systems, however, may be of higher dimensions). 
% The raw key (the key from which the secure key is established later) is extracted 
% from the pair $\{A0, B1\}$. 
the corresponding bit error rate is given by $ e = \mbox{prob} (a_0 \neq b_1)$, where \(a_0\) and \(b_1\) are the measurement results of \(A_0\) and \(B_1\). 
% However, we won't exploit the violation of Bell inequality, 
% an entanglement witness, to prove the security of  QKD. Instead, 
We will now use the complementarity between purity and quantum mutual information, as obtained above in Sec. \ref{pq-relations}, 
to establish the security of the protocol. Indeed, we 
will obtain a bound on the secret key rate by using the complementarity. 

A potential eavesdropper, Eve, denoted by \(E\), tries to gather information about the key of Alice and Bob. To this end, Eve plants ancillas near the channels carrying the 
states of Alice and Bob, and consequently, the Alice-Bob-Eve trio share the state \(\rho_{ABE}\). 

% From the perspective of QKD, consider the sender(Alice)-receiver(Bob)-eavesdropper(Eve) trio. Monogamy of quantum correlations ensures that if Alice and Bob are 
% sufficiently correlated then Alice (or Bob) and Eve are less correlated, i.e., the adversary Eve is essentially factorized out. Consequently, information can be 
% communicated securely. Initially, Alice-Bob and Alice-Eve are respectively maximally and least correlated when they share a quantum state of the form 
% $|\psi\rangle_{ABE}=|\phi^+\rangle_{AB}\otimes |\xi\rangle_E \in {\cal H}(\mathbb{C}^{d_{AB}}\otimes \mathbb{C}^{d_E})$. In the course of time, the adversary Eve 
% tries to build correlation with Alice and/or Bob and the resulting state can be written as $\rho_{ABE}=\sum_k A_k \rho^\psi_{ABE}A_k^{\dagger}$,
% %\begin{equation}
% %\rho_{ABE}=\sum_k A_k \rho^\psi_{ABE}A_k^{\dagger},
% %\end{equation} 
% where $A_k$ are Kraus operators satisfying $\sum_k A_k^{\dagger}A_k=I$. Its then natural to investigate the complementarity between purity of Alice-Bob 
% state $\rho_{AB}=\mbox{tr}_E (\rho_{ABE})$, and correlation ${\cal Q}'_{AB:E}\equiv {\cal Q}'(\rho_{AB:E})$ between Alice-Bob and Eve. Higher the purity 
% and lower the correlation ensures the security of communication.

{\it Key rates}.-- The optimal key rate for individual attacks and obtained via one-way communication  
%, \(r\), for the most general eavesdropping attack, 
%is bounded by the 
%one for individual attacks, and in the latter case, 
is provided by 
the Csisz\'{a}r-K\"{o}rner criterion \cite{ck-rate}, given by
\begin{equation}
%r \geq 
r_{CK} = \tilde{I}'_{A_0:B_1} - \mbox{min}\{\tilde{I}'_{A_0:\tilde{E}},\tilde{I}'_{B_1:\tilde{E}}\},
\end{equation} 
where we have used the notation \(\tilde{E}\)  for the measurement setting at Eve, and \(\tilde{I}'_{A_0:B_1}=1-h(e)\), with 
$h(p)=-p\log_2 p-(1-p)\log_2(1-p)$ being the binary entropy for \(0\leq p\leq 1\).
% For collective attacks, the secret key rate is given by the Devetak-Winter condition
% %under one-way classical post-processing from Bob to Alice is lower bounded by the Devetak-Winter rate 
% \cite{dw-rate}:
% \begin{equation}
% r \geq r_{DW}=\tilde{I}'_{A_0:B_1} - \mbox{min}\{\chi(E|A_0), \chi(E|B_1)\},
% \end{equation}
% where $\chi(E|{\cal X})=S(\rho_E) - \frac12 \sum_{x=\pm 1}S(\rho_{E|x})$ is the Holevo quantity \cite{Holevo} of Eve's ensemble given that the party \(X\) has measured 
% in the setting \({\cal X}\), with \(x=\pm 1\) denoting the measurement outcomes.
% %For the protocol in Ref. \cite{acin-protocol}, $\chi(E|A_0) \geq \chi(E|B_1)$.
% The symmetry of the protocol under consideration ensures that the probabilities of the outcomes at party \(X\) are equal. 

%Let us first focus on the case of individual attacks. 
Writing the complementarity between purity and quantum mutual information of \(\rho_{ABE}\) as 
\begin{equation}
 {\cal P}_{AB} + {\cal I}_{AB:E} \leq b,
\end{equation}
we have 
\begin{equation}
 \frac{I'_{AB:E}}{2 \min\{\log_2 d_{AB}, \log_2 d_E\}} \leq b - {\cal P}_{AB},
\end{equation}
following ineqs. (\ref{eq:qcr-qmi}) and (\ref{eq:qcr-qmi-abar}).
%The notational changes in the complementarity relation are the renaming of the party \(C\) as \(E\), and denoting the right-hand-side as \(b\). 
%
Now, quantum mutual information is non-decreasing under discarding of subsystems \cite{qmi-bunch},
so that 
\(I'_{A:E} \leq I'_{AB:E}\) and \(I'_{B:E} \leq I'_{AB:E}\). Therefore, 
\begin{equation}
 \min \{ I'_{A:E} ,  I'_{B:E}   \} \leq 2 \min\{\log_2 d_{AB}, \log_2 d_E\} (b - {\cal P}_{AB}).
\end{equation}
Furthermore, classical mutual information is upper bounded by the corresponding quantum mutual 
information \cite{molmer},  i.e. 
\(\tilde{I}'_{X:Y} \leq I'_{X:Y}\), resulting in 
\begin{equation}
 \min \{ \tilde{I}'_{A_0:\tilde{E}}, \tilde{I}'_{B_1:\tilde{E}} \} \leq 2 \min\{\log_2 d_{AB}, \log_2 d_E\} (b - {\cal P}_{AB}).
\end{equation}
Therefore, we have
\begin{equation}
 r_{CK} \geq 1 - h(e) - 2 \min\{\log_2 d_{AB}, \log_2 d_E\} (b - {\cal P}_{AB}).
\end{equation}
It is reasonable to allow the eavesdropper to be of a larger dimension, so that we choose \(d_{AB} \leq d_E\), whence
\begin{equation}
 r_{CK} \geq 1 - h(e) - 2 S(\rho_{AB}).
\end{equation}

% In the case of collective attacks, we again get the same bound:
% \begin{equation}
%  r \geq r_{DW} \geq 1 - h(e) - 2 S(\rho_{AB}),
% \end{equation}
% where we additionally require that the total state of the legitimate users of the key distribution protocol and the eavesdropper to be pure, again a reasonable 
% one in the general paradigm of ``the church of the larger Hilbert space'' \cite{church}.

To illustrate the rate obtained, consider that the legitimate users of the key distribution channel share 
the Werner state \cite{Bareilly-mey-bansh}, 
given by 
\(\rho_W = p|\phi^+\rangle \langle \phi^+| + (1-p) \frac{1}{2}I_2 \otimes \frac{1}{2}I_2\), where \(|\phi^+\rangle = \frac{1}{\sqrt{2}}(|00\rangle + |11\rangle)\), 
\(0\leq p \leq 1\), and 
\(I_2\) denotes the identity operator on the qubit Hilbert space. 
%Here \(|0\rangle\)
% and \(|1\rangle\) are the eigenvectors of the Pauli \(\sigma_z\) operator. 
In this case, \(e=\frac{1}{2}(1-p)\), and 
the entropy of the shared state between Alice and Bob is given by
\(H(\{\frac{1}{2}e,\frac{1}{2}e,\frac{1}{2}e,1-\frac{3}{2}e\})\), where \(H(\{p_i\})=-\sum_ip_i \log_2 p_i\) is the Shannon entropy of the probability distribution 
\(\{p_i\}\). Consequently, the maximal bit error rate that is allowed before the protocol becomes insecure is about \(3.6\)\%. The Shor-Preskill security proof 
provides a rate of \(11\)\% \cite{Shor-Preskill}. 

% It is interesting to compare the complementarity-based security proof here with the device independent security proofs of quantum key distribution 
% \cite{ekert91, acin-diqkd, diqkd-NJP}. 
Let us note here that if Alice and Bob 
share a pure state, then the quantum mutual information between 
the Alice-Bob system and the eavesdropper vanishes. It is therefore apparent that for large values of \(p\), 
in the shared Werner state, 
a stronger complementarity similar to 
\({\cal P}_{AB}+{\cal I}_{AB:E} \lesssim 1 \) is active, and this is independent of whether \(d_{AB} \leq d_E\) is valid. 
% Evidence for this also comes from the  numerical results on the complementarity between purity and 
% \(\min\{ \mathcal{Q}_{A:C}, \mathcal{Q}_{B:C} \}\).
Using this stronger complementarity would provide a security proof without the assumption \(d_{AB} \leq d_E\).
% bring down the allowed rate to below \(11\)\%, the maximal allowed rate from Shor and Preskill \cite{Shor-Preskill}. Moreover, it would provide a device independent 
% security proof, against collective attacks, via complementarity relations, instead of Bell inequality violations \cite{Bell}. 
We also note that there are recent results on device independent protocols  for estimating entropies \cite{orey-taruni}, 
which if possible to be used in the present scenario, would 
provide the potential for a device independent security proof of quantum cryptography \cite{acin-diqkd, diqkd-NJP} 
via information complementarity relations, instead of Bell inequality violations.
We mention here that the Devetak-Winter result \cite{dw-rate} also provides an entropic bound on the secret key rate for collective attacks. 
% also provides an entropic bound 
We however believe that the bound obtained here is via methods that provide an independent perspective on the Ekert protocol and its security, albeit for 
individual attacks.

\section{Conclusion}
\label{sec-conclu} 
To sum up, we have derived complementarity relations in  multiparty quantum systems connecting purity of a part of the system with 
correlations -- classical, quantum, and total -- between that part and the rest of the system.  We found that they have the potential to provide 
bounds on the secret key rates in quantum cryptography. 
%Subsequently, we discuss its connection to device independent quantum cryptography. 

%To our knowledge, the only 
A relation for three-party quantum states that has a similar topology to the ones derived here is in Ref. \cite{bristi-ekhono}, 
where local measurement-induced changes in two-party entanglement is related to the measurement-induced changes in 
 entanglement of those two parties with the third party, in a three party system. See also Ref. \cite{koashi-winter}.

Since the derived complementarity relations are rather unalike to the previously existing ones, they are potentially useful in getting 
 perspectives, hitherto not known, on protocols and 
phenomena involving many parties, including e.g. the black hole information paradox \cite{agami}.

%\section{Conclusion}
%\label{summary}

\begin{acknowledgments}
A.B. acknowledges an Inspire Fellowship,
%from the Department of Science and Technology, Government of India. 
and R.P. acknowledges an INSPIRE-faculty position at the Harish-Chandra Research Institute (HRI), from the Department of Science and Technology, Government of India. 
We acknowledge computations performed at the cluster computing facility at HRI.
\end{acknowledgments}


\begin{thebibliography}{99}
\bibitem{Garisto-Hardy} While maximally entangled mixed states are possible \cite{Garisto-Hardy-1}, the Ekert protocol will ultimately use one of its 
locally orthogonal components, and the argument goes through once again.
\end{thebibliography}
\end{document}